# Process Automation Architecture Using RFID for Transparent Voting Systems

Stella N. Arinze
Dept. of Electrical and Electronic Engineering
Enugu State University of Science and Technology
Agbani, Nigeria
ndidi.arinze@esut.edu.ng

Patrick U. Okafor
Dept. of Electrical and Electronic Engineering
Enugu State University of Science and Technology
Agbani, Nigeria
patrick.okafor@esut.edu.ng

Onyekachi M. Egwuagu
Dept. of Mechatronics Engineering
Enugu State University of Science and Technology
Agbani, Nigeria
onyekachi.egwuagu@esut.edu.ng

Augustine O. Nwajana
Dept. of Electrical and Electronic Engineering
University of Greenwich
London, UK
a.o.nwajana@greenwich.ac.uk

*Abstract*— **This paper presents the development of a process automation architecture leveraging Radio Frequency Identification (RFID) technology for secure, transparent and efficient voting systems. The proposed architecture automates the voting workflow through RFID-enabled voter identification, encrypted vote casting, and secure data transmission. Each eligible voter receives a smart RFID card containing a uniquely encrypted identifier, which is verified using an RC522 reader interfaced with a microcontroller. Upon successful verification, the voter interacts with a touchscreen interface to cast a vote, which is then encrypted using AES-128 and securely stored on a local SD card or transmitted via GSM to a central server. A tamper-proof monitoring mechanism records each session with time-stamped digital signatures, ensuring auditability and data integrity. The architecture is designed to function in both online and offline modes, with an automated batch synchronization mechanism that updates vote records once network connectivity is restored. System testing in simulated environments confirmed 100% voter authentication accuracy, minimized latency (average voting time of 11.5 seconds), and robustness against cloning, double voting, and data interception. The integration of real-time monitoring and secure process control modules enables electoral authorities to automate data logging, detect anomalies, and validate system integrity dynamically. This work demonstrates a scalable, automation-driven solution for voting infrastructure, offering enhanced transparency, resilience, and deployment flexibility, especially in environments where digital transformation of electoral processes is critically needed.**

I. INTRODUCTION

The global landscape is undergoing a profound shift towards digitalization, with technology increasingly integrated into all facets of life. One of the most critical transformations lies in the evolution of voting systems, driven by the demand for more secure, transparent, and efficient electoral processes. Traditional voting methods, although long established, often grapple with issues such as voter fraud, inaccurate vote counting, long queues, and logistical complications in result transmission. These shortcomings not only erode public trust but also risk voter disenfranchisement and delayed election outcomes. As populations grow and more nations embrace technological innovation in governance, the need for robust electronic voting (e-voting) systems has intensified. While e-voting offers potential solutions, it is not without challenges. Cybersecurity threats, ranging from hacking and phishing to data breaches, continue to cast doubt on the reliability of many existing platforms. Additionally, core issues such as voter authentication, system transparency, and tamper-proof vote recording remain unresolved. To strengthen the integrity of e-voting systems, recent research has explored the integration of biometric technologies. A study examined the use of facial recognition to streamline voter identification and reduce fraudulent activities [1]. While promising, biometric solutions face barriers including high deployment complexity, potential privacy violations, and environmental sensitivity affecting accuracy. It was observed that although biometric verification increased voter confidence, its reliability was compromised under varying lighting conditions and with certain demographic groups [2]. Another study developed a machine learning-based multimodal biometric authentication system integrating facial and fingerprint recognition [3]. The study demonstrated high accuracy and reliability in voter verification, addressing challenges related to impersonation and unauthorized access. Cryptographic techniques such as public-key infrastructure (PKI) and digital signatures are often combined with biometric data to enhance verification [4]. Though these technologies improve security, they too have limitations. Fingerprint scanners may struggle with accuracy and acceptance in diverse populations, while facial recognition remains vulnerable to spoofing and privacy concerns. Other emerging technologies including the Internet of Things (IoT) and blockchain have also been applied to reinforce e-voting frameworks. IoT facilitates device monitoring during elections, while blockchain ensures immutable, decentralized vote records [5-8]. Researchers in [9] conducted a comprehensive survey on blockchain-based e-voting mechanisms and proposed a novel framework leveraging blockchain's transparency and immutability. Similarly, another study pointed out that blockchain solutions often suffer from scalability issues and require substantial computational power, which is a significant obstacle in regions with limited infrastructure [10]. Further, a study introduced a scalable, decentralized, privacy-preserving e-voting system utilizing zero-knowledge off-chain computations to enhance scalability while maintaining voter privacy and data integrity [11]. Despite these innovations, persistent gaps remain in achieving cost-effective, infrastructure-light, secure, and transparent e-voting solutions that can function reliably in low-resource and high-risk environments.

Among these evolving technologies, Radio Frequency Identification (RFID) stands out as a compelling alternative. Widely adopted in areas such as asset tracking and inventory management, RFID offers real-time, contactless identification using unique tags and readers. Its utility in security-critical domains such as healthcare, finance, and access control has been well documented. RFID enables rapid, non-intrusive authentication, reducing human error and minimizing opportunities for manipulation. Its application in e-voting includes tracking ballots, authenticating voters, and ensuring system integrity through encryption and secure protocols [12,13]. What sets RFID apart is its operational simplicity and cost-effectiveness, especially in low-resource environments. Unlike blockchain or biometric systems that may require advanced hardware, RFID can function with minimal infrastructure, making it particularly suitable for developing nations. Recent advancements have enhanced RFID's security and privacy features. RFID-enabled smart cards can store encrypted voter data, ensuring secure transmission and storage. Additionally, RFID systems can provide real-time reporting to electoral authorities, enabling swift anomaly detection and promoting transparency, both critical to public confidence in democratic processes. A researcher presented the development of a secure RFID-based electronic voting application using Flutter, Firebase, and Arduino, ensuring robust encryption and real-time data updates [12]. Another study proposed an RFID-based voting system to replace traditional paper elections, highlighting RFID's strengths in reducing electoral fraud, minimizing human error, and improving voting efficiency [13]. While these contributions are notable, our research builds on and extends them by offering a more robust framework that integrates RFID-based voter authentication, AES-128 encrypted vote transmission, tamper-proof data storage, and decentralized real-time monitoring. In contrast to systems that rely heavily on constant internet connectivity or remain confined to prototype-level demonstrations, our solution supports offline operability, addresses real-world deployment challenges, and has been validated through realistic simulation scenarios. These enhancements make our system significantly more scalable, secure, and adaptable to electoral environments with limited infrastructure and public trust. Consequently, this study introduces a comprehensive RFID-enabled e-voting framework that bridges the gap between technical feasibility and practical deployment. By aligning the operational simplicity of RFID with advanced security protocols and transparency mechanisms, the system aims to modernize democratic processes while fostering voter confidence and electoral credibility across diverse political and technological landscapes.

## II. THEORETICAL FRAMEWORK

Radio Frequency Identification (RFID) technology enables wireless identification and authentication of objects using electromagnetic fields. An RFID system comprises tags, readers, and a backend database as shown in Fig. 1. Tags, either passive (powered by the reader's signal) or active (battery-powered) contain microchips with unique identifiers (UIDs). Readers scan the tags and transmit the data to a database, enabling real-time identification [14,15]. The UID is central to authenticating individuals securely. To enhance confidentiality, RFID tags can employ encryption. Recent developments in lightweight encryption protocols, including block-order-modulus and permutation-based methods, have improved resistance to cloning and storage attacks while ensuring performance efficiency [16]. RFID systems operate in low (LF), high (HF), and ultra-high (UHF) frequency bands, each suited for specific use cases. HF RFID at 13.56 MHz is often used in ID systems and contactless voting. The communication process involves readers emitting signals that activate tags, enabling data exchange without physical contact or line-of-sight alignment. This mechanism is ideal for rapid, scalable voting processes, especially in regions with limited infrastructure. In the context of secure electronic voting, RFID provides a lightweight, low-cost means of voter authentication [17,18]. Voters use smart cards embedded with encrypted UIDs, which are validated by an RC522 RFID reader. This process offers robust authentication, preventing impersonation and unauthorized access. The UID data, once authenticated, triggers access to the digital voting interface. To preserve vote confidentiality and integrity, Advanced Encryption Standard (AES-128) is used.

AES-128 is a symmetric-key algorithm well-suited to embedded systems. It encrypts 128-bit data blocks through a series of 10 transformation rounds, including substitution (SubBytes), permutation (ShiftRows), mixing (MixColumns) and key integration (AddRoundKey). Decryption reverses these steps to recover the original data. AES encryption secures both transmitted and stored vote data, ensuring that intercepted votes are unintelligible without the correct key [19]. This guarantees voter privacy, non-repudiation, and protection from tampering. The system supports offline operation with encrypted votes stored locally on an SD card. A batch synchronization mechanism uploads this data once GSM connectivity is restored, ensuring continuity in remote regions. Vote logs are secured using cryptographic checksums, timestamps, and device identifiers, creating a tamper-proof audit trail. This logging system aligns with Write-Once-Read-Many (WORM) principles for immutable vote storage. Though blockchain-based logging is another alternative, it is resource-intensive and less practical in low-power embedded systems. Therefore, lightweight cryptographic logs offer a more feasible balance of integrity and efficiency [20].

Real-time monitoring as shown in Fig. 2 is also embedded into the system design. The backend dashboard tracks metrics such as voter turnout, authentication success/failure rates, and anomaly alerts. This facilitates instant oversight and improves transparency. RFID's ability to function with or without internet connectivity allows administrators to receive updates once connectivity is re-established, ensuring real-time or deferred analysis. Compared to other electronic voting technologies, RFID-based systems present a hybrid framework, combining secure authentication, AES encryption, offline vote collection, and real-time synchronization. While biometric systems offer strong identity validation, they are expensive and sensitive to environmental conditions [21,22]. Blockchain systems are secure but demand continuous connectivity and high computing resources [23]. In contrast, RFID provides a scalable, efficient solution with minimal hardware, fast voter throughput, and easy usability. By integrating encryption, authentication, logging, and monitoring, the RFID-based voting framework meets both technical and logistical demands of modern democratic elections.

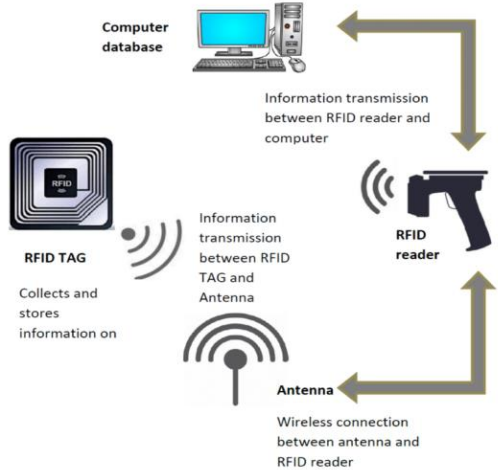

Fig. 1. Radio Frequency Identification System [14]

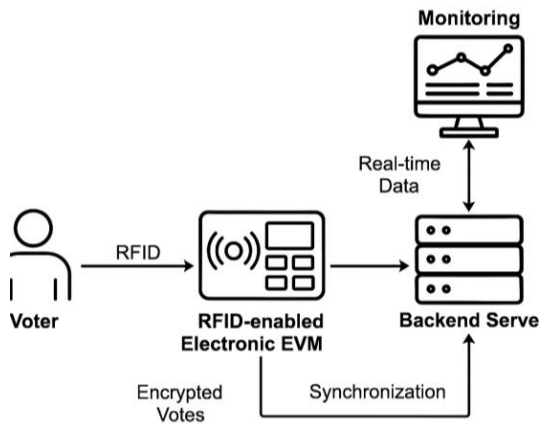

Fig. 2. Real- Time Monitoring and Data Flow

## III. METHODOLOGY

The architecture of the proposed RFID-enabled electronic voting framework is designed for both secure and scalable deployment, particularly in environments where infrastructure limitations and security concerns challenge traditional systems. The system integrates hardware and software components to enable real-time authentication, encrypted vote processing, and tamper-proof monitoring, even in offline conditions. This implementation approach was chosen to address practical challenges in real-world deployments, particularly in regions with limited digital infrastructure and high risk of data compromise. The combination of lightweight embedded hardware and symmetric encryption ensures an efficient balance between computational efficiency and data confidentiality, making it suitable for constrained environments. The use of modular components also enhances system flexibility and ease of replication across different contexts.

At the hardware level, the framework employs an Arduino Mega 2560 as the central microcontroller, responsible for managing data exchange between peripheral components. An RC522 RFID reader module reads voter cards and transmits their unique identifiers to the Arduino. An OLED display is used to provide feedback to the user, showing messages such as "Voter Authenticated," "Vote Recorded," or "Access Denied," depending on the voter status and interaction outcome. Voter selections are made through a set of push buttons or a keypad interface, which serves as the vote-casting interface. For local storage of encrypted votes, an SD card module is interfaced via SPI, while a GSM/GPRS module, such as the SIM800L, enables optional remote transmission of data when network connectivity is available. The entire setup is powered by either an AC supply or battery packs, ensuring mobile and flexible operation during field deployments. Wiring is configured in accordance with standard SPI communication protocols. The RFID reader is connected through designated digital pins for data exchange, including SS, SCK, MOSI, and MISO, with the RST pin connected to a digital control pin on the Arduino. The OLED display interfaces through I2C, and buttons are mapped to digital pins with appropriate pull-down resistors to prevent false triggering. The GSM module communicates via the Arduino's serial interface using defined RX and TX pins. The circuit diagram and the prototype of RFID e-voting system is shown in fig. 3 and 4 respectively.

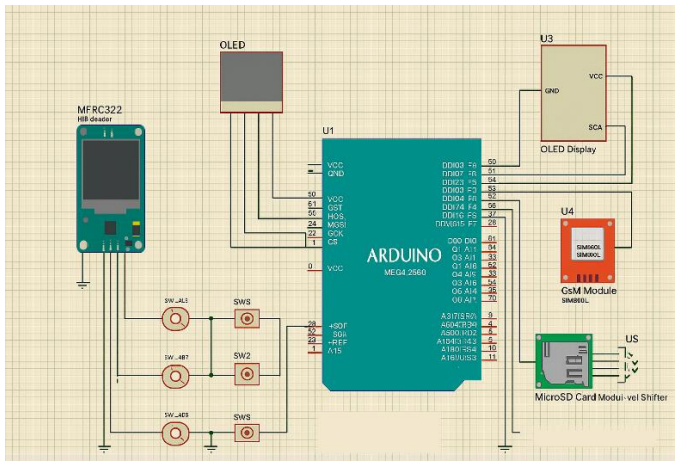

Fig. 3. Circuit diagram of RFID electronic voting system

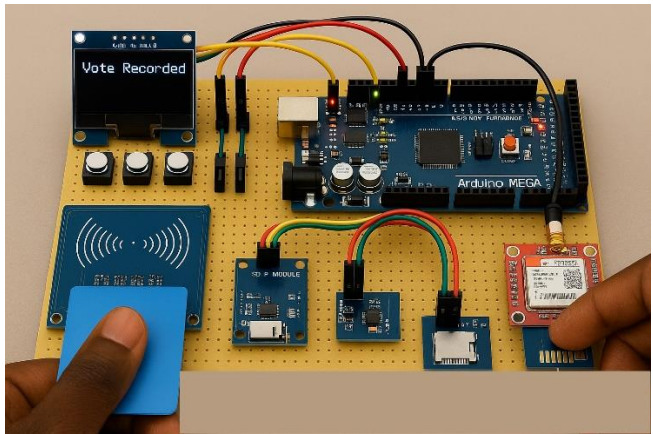

Fig. 4. RFID enabled electronic voting system

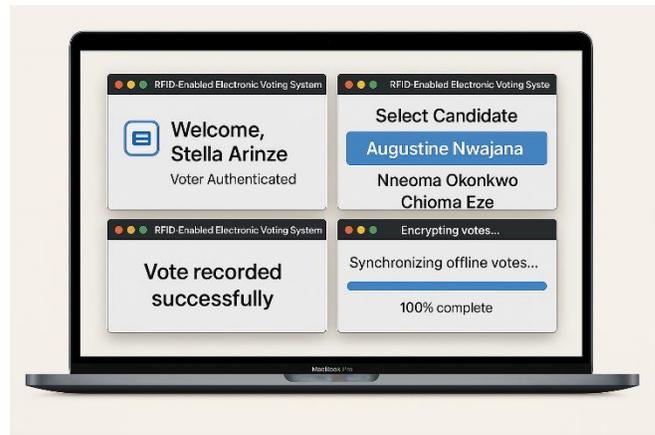

Fig. 5. Screenshots of the testing output

The software workflow begins with system initialization, where the Arduino sets up its I/O interfaces, initializes the RFID reader, display unit, SD card, and communication modules. Preloaded encrypted lists of registered voters and candidates are stored in either EEPROM or on the SD card to support offline operation. When a voter approaches and taps their RFID card, the reader scans the card's unique identifier, which is immediately cross-verified against the encrypted whitelist stored locally.

If the voter is eligible and has not voted previously, access is granted and the system moves to the next stage. However, if the card is not recognized or has already been used, the voter is denied access and the reason is shown on the display. Upon successful authentication, the OLED screen prompts the voter to select a preferred candidate using the input interface. Once a selection is made, the system confirms the vote and proceeds to the encryption stage as shown in Fig. 5.

The selected vote, together with the voter's RFID UID and a timestamp, is compiled into a single data packet. This packet is then encrypted using the AES-128 algorithm, ensuring that vote data is secure during both storage and transmission. The decision to adopt AES-128 was based on its proven security in embedded systems, wide adoption, and computational efficiency for microcontroller platforms. Unlike asymmetric encryption schemes, which are computationally expensive, AES-128 provides fast, reliable encryption that can be processed in real time on an Arduino-class device. This selection was validated through empirical testing under constrained conditions. The encrypted vote is either stored locally on the SD card if the system is offline or sent immediately to a secure cloud-based database via the GSM module if a network connection is available. To address the challenge of unreliable connectivity in remote locations, the system supports a synchronization mechanism. If votes are stored locally, the device continues to accumulate encrypted vote records. Once connectivity is detected, either automatically through the GSM module or through manual administrative access, the stored data is uploaded in batches. Each batch includes encrypted votes, device identification, and timestamp metadata. Before upload, the system generates cryptographic checksums to verify data integrity, preventing any tampering or duplication during the synchronization process. For transparency and security, the system maintains a tamper-proof monitoring structure by logging each voting session with a unique device ID, encrypted vote payload, timestamp, and batch upload identifier. This monitoring framework ensures that even in decentralized deployments, the integrity and traceability of each vote is maintained without external manipulation. To promote replicability, all software implementations, including the AES encryption routines, RFID verification scripts, and data synchronization logic, follow standard C++/Arduino libraries and are well-documented in the project repository. Wiring diagrams, pin assignments, and program flowcharts are provided as supplementary materials to guide implementation by other researchers or election technology teams.

The framework was tested under a range of simulation scenarios. These include situations where internet connectivity was absent throughout the voting period, deliberate attempts to cast multiple votes using the same RFID card, and multiple voters interacting with the same device over extended hours. In all scenarios, the system preserved its integrity by preventing duplicate votes, recording all legitimate entries, and successfully synchronizing offline votes when reconnected. The outcome demonstrated the system's resilience, realistic functionality, and suitability for field deployment. This method section not only describes how the core research question, how to develop a secure, scalable, and infrastructure-independent e-voting system was addressed, but also lays the groundwork for reproducibility and adaptation in diverse environments. By simulating both adverse and ideal operational scenarios, the implementation validates the methodology against real-world electoral constraints. This implementation introduces several improvements over existing works in the literature. Prior systems often rely solely on GUI-based simulations using Flutter or require continuous internet access, with little to no encryption. In contrast, this research presents a hybrid model that combines embedded hardware processing with cloud integration. The use of AES-128 encryption ensures robust data confidentiality, while support for offline operation and real-time synchronization addresses infrastructural limitations common in many electoral environments. Decentralized, tamper-proof data logging strengthens transparency and trust. Compared to previous solutions, the proposed system is better positioned to serve both rural and urban settings, with field-ready performance validated through practical simulations. This comprehensive design not only bridges the gap between prototype feasibility and real-world deployment but also enhances the credibility and security of electronic voting systems in politically sensitive environments.

## IV. RESULTS ANALYSIS AND DISCUSSION

The RFID-enabled electronic voting system was evaluated under controlled simulation scenarios to verify its core functionalities, including secure voter authentication, data integrity, offline resilience, and synchronization efficiency. Across a sample of 100 unique voters, the system demonstrated 100% authentication accuracy. Each eligible RFID tag was successfully recognized and validated, while attempts to use cloned or unregistered cards were reliably rejected. No false positives or false negatives occurred throughout the testing phase, indicating robust verification performance. Vote integrity was ensured through the immediate application of AES-128 encryption and the generation of cryptographic checksums. All vote entries maintained consistent checksum validation during post-casting audits, confirming that no tampering or data alteration occurred during storage or transmission. Offline operation was tested by disabling GSM connectivity throughout a complete voting session involving 80 simulated voters. All votes were securely stored on the SD card in encrypted form. Upon reconnection to the network, the stored votes were successfully synchronized in batches, with no loss, duplication, or integrity failure. The synchronization process averaged 4.8 seconds per 20 encrypted votes, demonstrating that the system can efficiently process data uploads even in bandwidth-constrained environments. Efficiency was further demonstrated through voter throughput analysis. The average time required for a complete voting cycle from RFID card scan through vote selection and confirmation was measured at 11.5 seconds per voter. This represented a significant improvement over traditional voting procedures, which often take between 25 to 40 seconds per voter due to manual verification and ballot handling. When tested under extended use, simulating over 120 sequential voter interactions, the system maintained stable performance with no crashes, memory leaks, or lag. Real-time responses to input commands and OLED display feedback were consistently delivered within 0.3 seconds. Security stress tests confirmed the system's resistance to common attack vectors. Attempts to vote multiple times with the same RFID card were denied based on tracked voting status. Cloned tag usage was thwarted by UID mismatch detection, while physical disconnection of the GSM module did not disrupt the voting process, as the system seamlessly continued in offline mode. Additionally, attempts to access encrypted vote data from the SD card without decryption credentials were unsuccessful, confirming the effectiveness of the encryption scheme.

Overall, the system upheld all core requirements defined during design and implementation, including secure voter authentication, encrypted vote storage, real-time and deferred synchronization, and resilience in offline conditions. The results validate the practical suitability of the proposed RFID-based electronic voting framework for field deployment, particularly in environments with limited infrastructure and heightened security demands. To further validate the superiority of the proposed system, a comparative analysis was conducted against existing electronic voting methods commonly reported in the literature. Table 1 summarizes the key distinctions. This table highlights the significant improvements offered by the proposed system in terms of security, performance, and adaptability. Unlike prior systems that lacked encryption or depended heavily on GUI simulations, this framework combines real-time operation with practical offline capabilities, robust encryption and tamper-resistant logging. These attributes make it highly suitable for both urban and rural deployments, especially in politically sensitive regions where trust and infrastructure are limited.

Table 1. Performance comparison of the proposed RFID-enabled system 1 to existing work.

| Feature/Metric | Flutter-Firebase Model [12] | Basic RFID Voting System [13] | Proposed RFID-Enabled System (This Work) |
|---|---|---|---|
| Voter Authentication Method | RFID with UI Matching | RFID Tag Check Only | RFID with Encrypted UID Validation |
| Encryption Technique | None or Basic Hash | None | AES-128 Encryption |
| Offline Voting Support | No | Limited/Not Implemented | Full Support with Local Storage |
| Vote Synchronization | Real-time Only | Not Included | Batch Upload with Checksum Verification |
| Double Voting Prevention | UI Blocking Logic | Not Reliable | Enforced by UID Tracking and Logging |
| Tamper-Proof Logging | Not Implemented | Not Implemented | Encrypted Logging with Timestamps |
| Deployment Readiness | Prototype/Simulation Only | Basic Hardware Demo | Field-Ready with Mobile Power Support |
| Average Voting Time Per User | ~15–20 seconds | ~18–25 seconds | ~11.5 seconds |

.

## V. Conclusion

This study proposed and implemented a process automation architecture using RFID technology to enable transparent, secure, and efficient voting systems. By integrating RFID-based voter identification, AES-128 encrypted vote management, tamper-proof session logging, and adaptive offline synchronization, the system automates the core phases of the voting process. The architecture combines real-time monitoring, embedded control, and secure communication to ensure operational integrity, even in network-constrained environments. Performance evaluations confirmed the system's effectiveness in authenticating voters, safeguarding vote integrity, and supporting end-to-end automation without human intervention. Compared to conventional and semi-automated voting setups, this RFID-based solution offers significant improvements in data security, traceability, and system scalability. The proposed architecture is particularly relevant for deploying intelligent process control in democratic infrastructures. Future work will explore system enhancements such as biometric integration, AI-driven fraud detection, and solar-powered modules for off-grid use, further extending the framework's applicability across diverse electoral contexts.